\documentclass[twocolumn,showpacs,preprintnumbers,amsmath,amssymb]{revtex4}

\usepackage{graphicx}
\usepackage{dcolumn}
\usepackage{bm}

\newcommand{\ie}{i.e.,}
\newcommand{\eg}{e.g.,}
\newcommand{\exc}{_{\rm ex}}
\renewcommand{\a}{^{(a)}}
\newcommand{\parent}{\rho^{(0)}}

\begin{document}

\title{Equilibrium phase behavior of polydisperse hard spheres}

\author{Moreno Fasolo}
 \email{moreno.fasolo@kcl.ac.uk}
\author{Peter Sollich}
 \email{peter.sollich@kcl.ac.uk}
\affiliation{
Department of Mathematics, King's College London, London WC2R 2LS, U.K.
}

\date{\today}
\begin{abstract}
We calculate the phase behavior of hard spheres with size
polydispersity, using accurate free energy expressions for the fluid
and solid phases.  Cloud and shadow curves, which determine the onset
of phase coexistence, are found exactly by the moment free energy
method, but we also compute the complete phase diagram, taking full
account of fractionation effects. In contrast to earlier, simplified
treatments we find no point of equal concentration between fluid and
solid or re-entrant melting at higher densities.  Rather, the fluid
cloud curve continues to the largest polydispersity that we study
(14\%); from the equilibrium phase behavior a terminal polydispersity
can thus only be defined for the solid, where we find it to be around
7\%.  At sufficiently large polydispersity, fractionation into several
solid phases can occur, consistent with previous approximate
calculations; we find in addition that coexistence of several solids
with a fluid phase is also possible.
\end{abstract}



\maketitle


During the past few decades, a great deal of effort has been devoted
to studies of the phase behavior of spherical particles, and in
particular of the freezing transition, where the particles arrange
themselves into a crystal with long-range translational order. The
simplest system for studying this transition is one where the
particles act as hard spheres, exhibiting no interaction except for an
infinite repulsion on overlap. This scenario can be realized
experimentally, using \eg\ colloidal latex particles sterically
stabilized by a polymer coating~\cite{PusVan86}. Hard spheres
constitute a purely entropic system; the internal energy $U$ vanishes,
and $F=-TS$. Phase transitions are thus entropically driven;
nevertheless, monodisperse (\ie\ identically sized) hard spheres
exhibit a freezing transition, where a fluid with a volume fraction of
$\phi\approx 50$\% coexists with a crystalline solid with $\phi\approx
55$\%~\cite{Pusey91}.

For colloidal hard spheres, there is inevitably a spread in the
particle diameters $\sigma$, which are effectively continuously
distributed within some interval. The width of the diameter
distribution can be characterized by a polydispersity parameter
$\delta$, defined as the standard deviation of the size distribution
divided by its mean.


The effect of polydispersity on the phase behavior of hard spheres
has been investigated by experiments~\cite{PusVan86,Pusey91}, computer
simulations~\cite{DicPar85,BolKof96,BolKof96b,PhaRusZhuCha98}, density
functional theories~\cite{BarHan86,McrHay88}, and simplified
analytical
theories~\cite{Pusey87,PhaRusZhuCha98,Bartlett97,Bartlett98,Sear98,BarWar99,XuBau03};
Ref.~\cite{PhaRusZhuCha98} has a more detailed bibliography of earlier
work. These studies have revealed that, compared to the monodisperse
case, polydispersity causes several qualitatively new phenomena.
First, it is intuitively clear~\cite{Pusey87} that significant
diameter polydispersity should destabilize the crystal phase, because
it is difficult to accommodate a range of diameters in a lattice
structure.  Experiments indeed show that crystallization is suppressed
above a {\em terminal polydispersity} of $\delta_t \approx
12\%$~\cite{PusVan86,Pusey91}. Theoretical work suggested that this
arises from a progressive narrowing of the fluid-solid coexistence
region with increasing $\delta$, with the phase boundaries meeting at
$\delta_t$~\cite{McrHay88,Bartlett97} in a point of equal
concentration~\cite{BarWar99}. Bartlett and Warren~\cite{BarWar99}
also found {\em re-entrant melting} on the high-density side of this
point: for $\delta$ just below $\delta_t$, they predicted that
compressing a crystal could transform it back into a fluid, as
sketched in the inset of Fig.~\ref{fig:cloud_shadow} below.
However, none of these theoretical studies fully accounted for {\em
fractionation}~\cite{Sollich02}, \ie\ the fact that coexisting phases
generally have different diameter distributions; in fluid-solid
coexistence, one typically finds that the solid contains a higher
proportion of the larger particles.
Beyond the resulting difference in mean diameter, fractionation
implies that coexisting phases can also have different
polydispersities $\delta$. Indeed, numerical simulations that allow
for fractionation show that a solid with a narrow size distribution
can coexist with an essentially arbitrarily polydisperse
fluid~\cite{BolKof96b,KofBol99}, suggesting that the concept of a
terminal polydispersity is useful only for the solid but not for the
fluid.
Fractionation has also been predicted to lead to {\em solid-solid
coexistence}~\cite{Bartlett98,Sear98}, where a broad diameter
distribution is split into a number of narrower solid fractions. This
occurs because the loss of entropy of mixing is outweighed by the
better packing, and therefore higher entropy, of crystals with narrow
size distribution; accordingly, as the overall polydispersity of the
system grows, the number of coexisting solids is predicted to
increase.

Previous work as described above leaves open a number of questions.
The drastic and differing approximations for size fractionation used
in the studies of re-entrant melting and solid-solid
coexistence~\cite{BarWar99,Bartlett98,Sear98} leave the relative
importance of these two phenomena unclear. In~\cite{BarWar99}
fractionation was allowed, but coexisting phases were implicitly
constrained to have the same $\delta$; calculations that account fully
for fractionation remain restricted to highly simplified van der Waals
free energies~\cite{XuBau03}. Numerical simulations have
been carried out at constant chemical potential
distribution~\cite{BolKof96,KofBol99}; in contrast to the experimental
situation, the overall particle size distribution can then change
dramatically across the phase diagram, limiting the applicability of
the results.

Our goal in this letter is to calculate the equilibrium phase
behavior of polydisperse hard spheres on the basis of accurate free
energy expressions, taking full account of fractionation and going beyond
previous work on fluid-solid and solid-solid coexistence.
The experimentally observed behavior of hard sphere colloids will of
course also depend on non-equilibrium effects, \eg\ the presence of a
kinetic glass transition~\cite{PusVan87}, anomalously large nucleation
barriers~\cite{AueFre01} or the growth kinetics of polydisperse
crystals~\cite{EvaHol01}. Nevertheless, the equilibrium phase
behavior needs to be understood as a baseline from which
non-equilibrium effects can be properly attributed. Also, more of the
equilibrium behavior may be observable under microgravity conditions,
where the glass transition is shifted to higher densities or even
absent~\cite{ZhuLiRogMeyOttRusCha97}.


Our calculations will show that the fluid cloud curve, which locates
the onset of phase coexistence coming from low density, continues to
large polydispersities $\delta$: the point of equal concentration
found in~\cite{BarWar99} disappears together with the predicted
re-entrant melting. Instead of returning to a single-phase fluid at
high volume fractions, the system splits into two or more fractionated
solids, consistent with the simplified calculations
of~\cite{Bartlett98}; coexistence of several solids with a fluid phase
appears as a new feature.

In general, the total free energy (density) of a polydisperse system
consists of an ideal and an excess part ($f\exc$). In units where
$k_{\rm B}T=1$, 
\begin{equation}
f = \int d\sigma \rho(\sigma) [\ln\rho(\sigma) -1] + f\exc
\label{eq:free_energy}
\end{equation}
Here $\rho(\sigma)$ is the density distribution, \ie\
$\rho(\sigma)\,d\sigma$ is the number density of particles with
diameters between $\sigma$ and $\sigma+d\sigma$. Equilibrium requires
equality of the chemical potentials $\mu(\sigma)=\delta f/\delta
\rho(\sigma)$ and of the pressure $\Pi = -f + \int\!d\sigma
\mu(\sigma)\rho(\sigma)$ among all coexisting phases $a=1\ldots
P$. Particle conservation adds the condition that, if phase $a$
occupies a fraction $v\a$ of the system volume, then $\sum_a
v\a\rho\a(\sigma)=\parent(\sigma)$, where $\parent(\sigma)$ is the
overall or ``parent'' density distribution.

For the {\em fluid}, the most accurate free energy approximation
available at present is the BMCSL
generalization~\cite{Boublik70,ManCarStaLel71} of the monodisperse
Carnahan-Starling equation of state. This is {\em truncatable} in the
sense that the excess free energy only depends on the four {\em
moments} $\rho_i=\int\!d\sigma\,\sigma^i\rho(\sigma)$ ($i=0\ldots 3$)
of the density distribution~\cite{SalSte82}; $\rho_0$ is the total
number density, $(\pi/6)\rho_3=\phi$ the volume fraction, and
$\rho_1/\rho_0=\bar{\sigma}$ and $\rho_2/\rho_0=\overline{\sigma^2}$
give the mean and mean-square diameter. For the crystalline {\em
solid}, Bartlett~\cite{Bartlett97,Bartlett99} assumed that the same
truncatable structure holds; an approximate excess free energy
(depending only on the same $\rho_i$) can then be derived from
simulation results~\cite{KraFre91} for bidisperse hard spheres.
Implicit in the use of data from~\cite{KraFre91} is the assumption
that the crystal has a substitutionally disordered f.c.c.\ structure.

We adopt the BMCSL and Bartlett free energies for our calculation; the
appropriate branch for a given $\rho(\sigma)$ is selected by taking
the minimum of the fluid and solid free energies. Since the excess
free energies depend only on the $\rho_i$, the excess chemical potentials
$\mu\exc(\sigma)$ take the form
\begin{equation}
\mu\exc(\sigma) = {\delta f\exc}/{\delta \rho(\sigma)} = 
\mu_0 + \mu_1 \sigma + \mu_2 \sigma^2 + \mu_3
\sigma^3
\label{eq:ex_che_pot}
\end{equation}
For the solid, Bartlett~\cite{Bartlett99} derived $\mu_0$ and $\mu_3$
from the small and large $\sigma$ limits of the Widom insertion
principle~\cite{Widom63}. However, because of the approximate
character of the excess free energy, $\mu\exc(\sigma)$ then does not
obey the thermodynamic consistency requirement
$\delta\mu\exc(\sigma)/\delta\rho(\sigma') =
\delta\mu\exc(\sigma')/\delta\rho(\sigma)$. To avoid this, we assign
all excess chemical potentials by explicitly carrying out the
differentiation in~(\ref{eq:ex_che_pot}).


Our computational approach is based on the moment free energy
method~\cite{SolCat98,Warren98,SolWarCat01}, which maps the full free
energy~(\ref{eq:free_energy}), with its dependence on all details of
$\rho(\sigma)$ through the ideal part, onto a moment free energy
depending only on the moments $\rho_i$. For truncatable free energies
this locates exactly the cloud points, \ie\ the onset of phase
separation coming from either a single-phase fluid or solid, as well
as the properties of the coexisting ``shadow'' phases that appear
there. Inside the coexistence region, one in principle needs to solve
a set of highly coupled nonlinear equations~\cite{Sollich02} and the
moment free energy method gives only approximate results.
However, by retaining extra moments with adaptively chosen weight
functions~\cite{ClaCueSeaSolSpe00,SolWarCat01,SpeSol03a}, increasingly
accurate solutions can be obtained by iteration. Using these as initial
points, we are then able to find full solutions of the exact phase
equilibrium equations. Care is taken to check that solutions are
globally stable, \ie\ that no phase split of lower free energy
exists~\cite{SolWarCat01}. We are able to calculate coexistence of up
to $P=5$ phases, which so far has been possible only for much simpler
free energies depending on a single density moment
(see \eg~\cite{SolWarCat01}).

Below we present results for a symmetric triangular parent density
distribution, \ie\ $\parent(\sigma)$ increasing linearly from zero for
$\sigma\in[1-w,1]$ and decreasing linearly for $\sigma\in[1,1+w]$,
with $w=\sqrt{6}\delta$. The mean diameter of 1 fixes our length
unit. Other distributions could be considered, but for the moderate
values of $\delta$ of interest here one expects them to give
qualitatively similar results, based on the intuition that for narrow
size distributions $\delta$ is the key parameter controlling the phase
behavior~\cite{Pusey87}.

\begin{figure}[h]
\includegraphics[width=8.5cm]{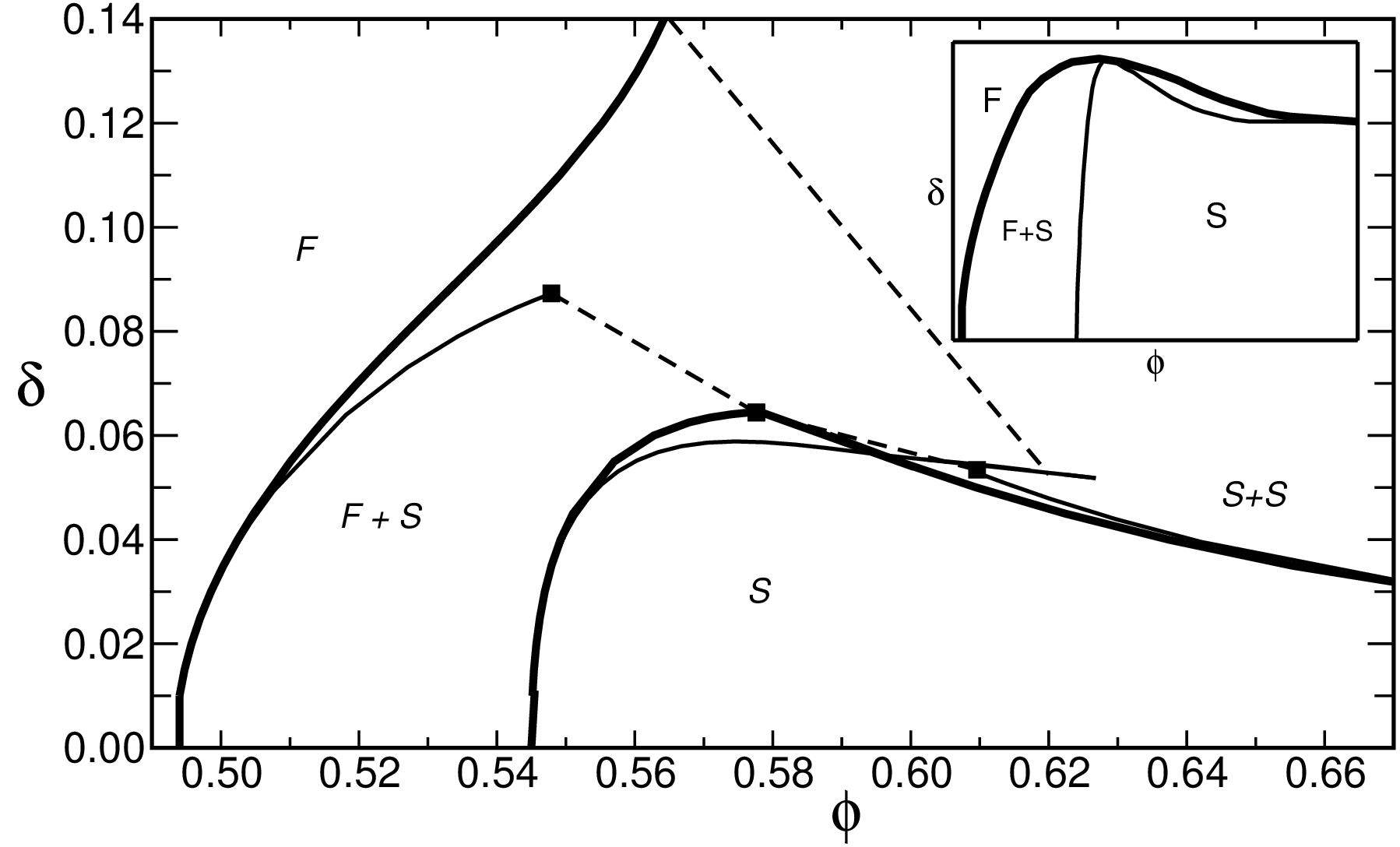}%
\caption{Cloud (thick) and shadow (thin) curves,
plotted as polydispersity $\delta$ versus volume fraction $\phi$;
dashed lines link sample cloud-shadow pairs.  The fluid (F) cloud
curve continues up to the largest $\delta$ that we study.
The solid (S) cloud curve has two branches, with onset of F-S and S-S
coexistence at low and high volume fractions, respectively. Inset:
Sketch of the phase diagram of~\cite{BarWar99}, showing re-entrant
melting and the point of equal concentration.
\label{fig:cloud_shadow}
}
\end{figure}

Fig.~\ref{fig:cloud_shadow} shows our results for the cloud and shadow
curves. The fluid cloud curve continues throughout the whole range of
polydispersities that we can investigate: even at $\delta=14\%$, a
hard sphere fluid will eventually split off a solid on compression.
Fractionation is key here; as indicated in
Fig.~\ref{fig:cloud_shadow}, the coexisting shadow solid always has a
smaller polydispersity, with $\delta$ never rising above $6\%$.  This
fractionation effect prevents the convergence of the solid and fluid
phase boundaries, along with the resulting re-entrant
melting~\cite{BarWar99} (Fig.~\ref{fig:cloud_shadow}-inset). These
findings are in qualitative accord with numerical simulations for the
simpler case of fixed chemical
potentials~\cite{BolKof96b,KofBol99}. In particular, the terminal
polydispersity $\delta_t$ cannot be defined as the point beyond which
a fluid at equilibrium will no longer phase separate; $\delta_t$ only
makes sense as the maximum polydispersity at which a single solid phase can
exist. As in~\cite{BolKof96b} we also find that the coexisting fluid
always has a lower volume fraction than the solid, along with (not
shown) a lower mean diameter.

\begin{figure}[b]
\includegraphics[width=8.5cm]{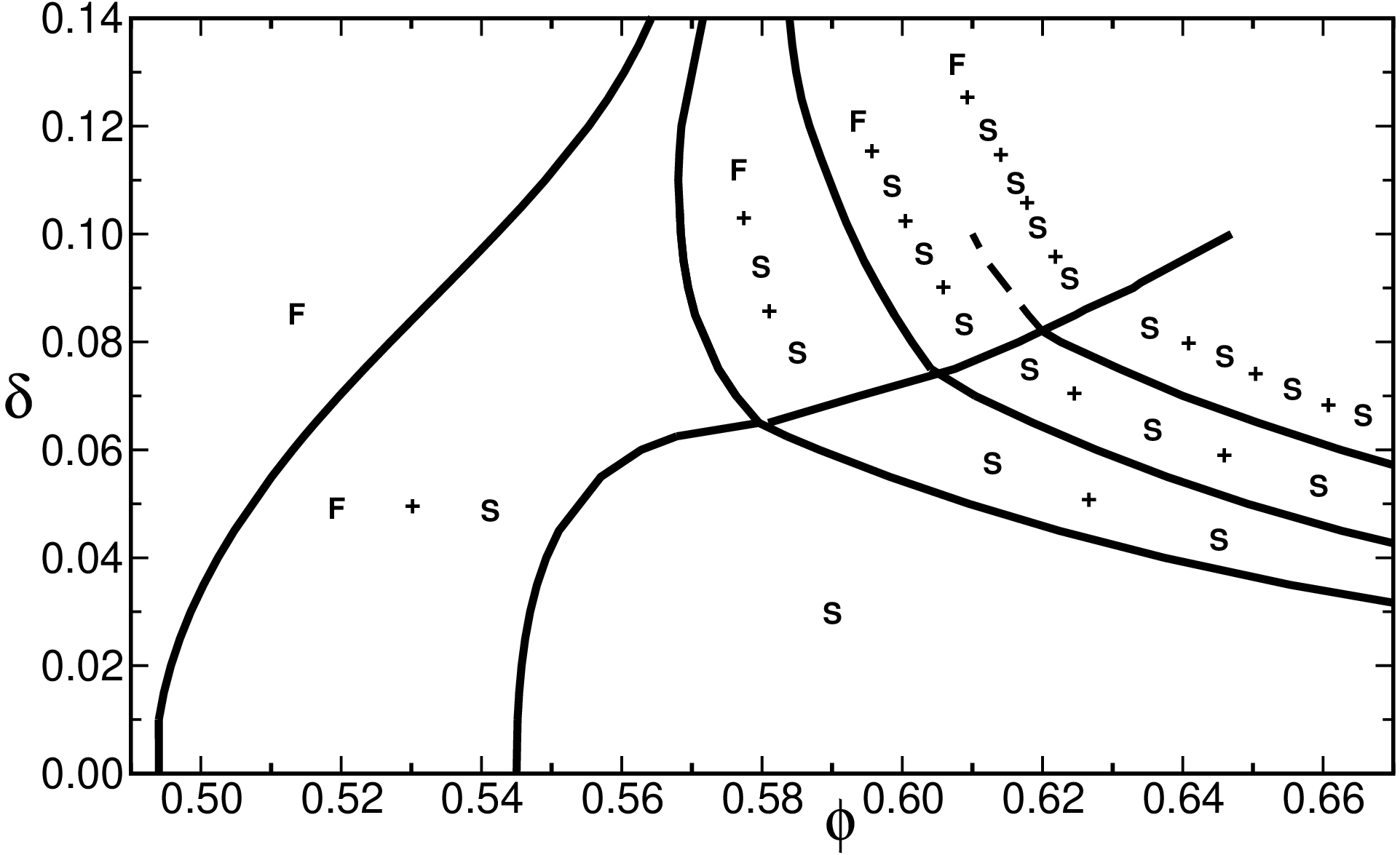}%
\caption{Full phase diagram for polydisperse hard spheres with a
triangular size distribution. In each region the nature of the
phase(s) coexisting at equilibrium is indicated (F: fluid, S:
solid). Dashed line: best guess for the phase boundary
in the region where our numerical data become unreliable.
\label{fig:parent}
}
\end{figure}

Coming from the single-phase solid, decreasing density at low
polydispersities leads to conventional fluid-solid phase
separation. At higher $\delta$, however, the solid cloud curve
acquires a second branch at higher densities. 
This is broadly analogous to the re-entrant phase boundary found
in~\cite{BarWar99}, but with the crucial difference that the system
phase separates into two solids rather than a solid and a fluid.
The two branches meet at a
triple point. Here the solid cloud phase coexists with two shadow
phases, one fluid and one solid, as marked by the squares in
Fig.~\ref{fig:cloud_shadow}. From Fig.~\ref{fig:cloud_shadow} the
triple point, at $\delta_t\approx 7\%$, also gives the terminal
polydispersity beyond which solids with triangular diameter
distribution are unstable against phase separation. As explained,
other distributions should give similar values of $\delta_t$.

\begin{figure}[h]
\includegraphics[width=8.5cm]{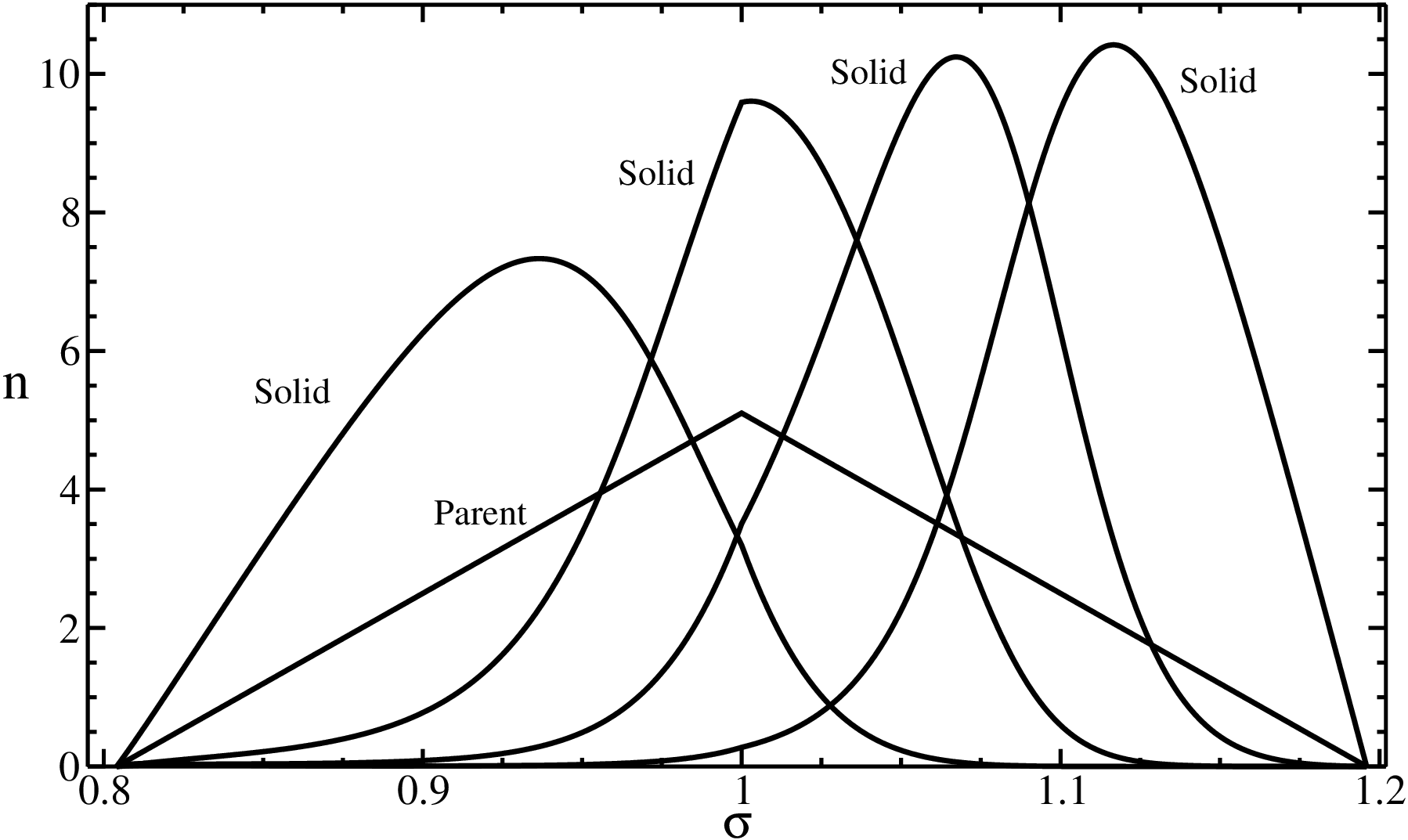}%
\caption{Normalized diameter distribution
of four coexisting solid phases obtained
from a parent with $(\phi,\delta)=(63.0\%,8\%)$. From left to
right, the solids
have volume fractions and polydispersities $(60.1\%,5.4\%)$, $(62.9\%,
4.6\%)$, $(64.6\%, 4.0\%)$, $(66.3\%, 3.6\%)$. 
\label{fig:solid_distribution}
}
\end{figure}

In Fig.~\ref{fig:parent} we show the full phase diagram for our
triangular parent distribution. In each region the nature of the
phase(s) coexisting at equilibrium is indicated. The cloud curves of
Fig.~\ref{fig:cloud_shadow} reappear as the boundaries between
single-phase regions and areas of fluid-solid or solid-solid
coexistence. Starting from the latter and increasing density or
$\delta$, fractionation into multiple solids occurs. The overall shape
of the phase boundaries in this region is in good qualitative
agreement with the approximate calculations
of~\cite{Bartlett98}. However, the coexisting solids do not
necessarily split the diameter range evenly among themselves as
assumed in~\cite{Bartlett98}; see the sample plot in
Fig.~\ref{fig:solid_distribution} of the normalized diameter
distributions $n(\sigma)=\rho(\sigma)/\rho_0$ of four coexisting
solids. In fact, plotting $\delta$ vs $\phi$ for all coexisting solids
across the phase diagram, we find points that cluster very closely
around the high-density branch of the solid cloud curve in
Fig.~\ref{fig:cloud_shadow}. Coexisting solids with lower volume
fraction $\phi$ thus tend to have higher polydispersity $\delta$, as
in the example in Fig.~\ref{fig:solid_distribution}; this conclusion
is intuitively appealing since higher compression should disfavor a
polydisperse crystalline packing.

Note that in Fig.~\ref{fig:parent}, at larger $\delta$ than we can
tackle numerically, coexistence of $P>4$ solids would be expected
since each individual solid can only tolerate a finite amount of
polydispersity. However, from Fig.~\ref{fig:parent} such phase splits
would occur at increasing densities and eventually be limited by the
physical maximum volume fraction $\phi_{\rm max}\approx 74\%$. Also,
at higher $\delta$ more complicated single-phase crystal structures,
with different lattice sites occupied preferentially by
(say) smaller and larger spheres, could appear and compete with the
substitutionally disordered solids we consider.

\begin{figure}
\includegraphics[width=8.5cm]{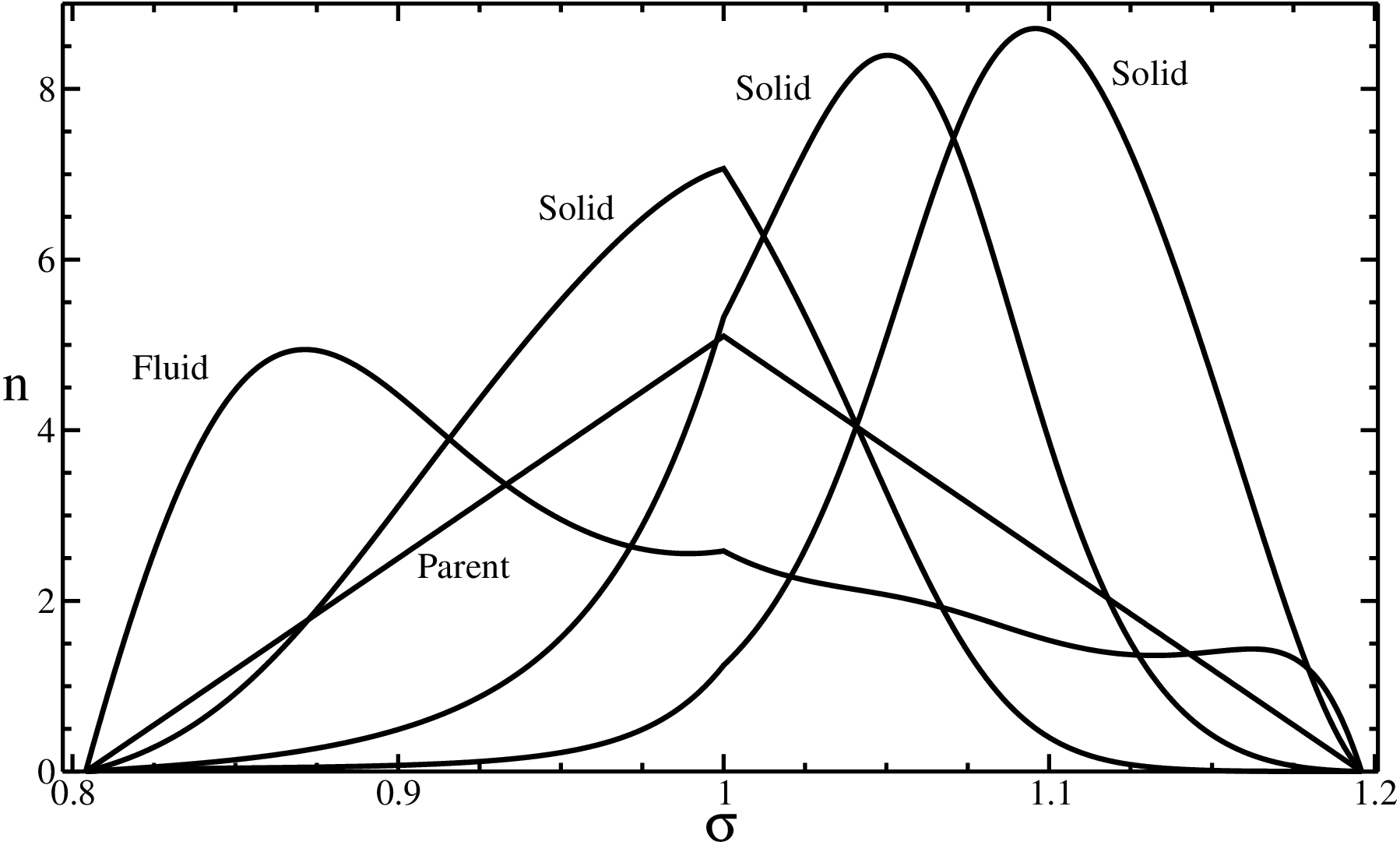}%
\caption{Normalized diameter distributions for F-S-S-S phase
coexistence obtained from a parent with $(\phi, \delta)=(60.3\%,
8\%)$.
\label{fig:distribution}
}
\end{figure}

Finally, a new feature of the phase diagram in Fig.~\ref{fig:parent}
is the coexistence of a fluid with multiple solids. The triple point
on the solid cloud curves already indicated the existence of a
three-phase F-S-S region; as in the case of solid-solid phase splits,
more solid phases then appear with increasing
$\delta$. Fig.~\ref{fig:distribution} shows again that the
fractionation behavior is non-trivial: while the coexisting fluid is
enriched in the smaller particles as expected, it also contains ``left
over'' large spheres that did not fit comfortably into the solid
phases and thus ends up having a {\em larger} polydispersity (10.4\%) than
the parent (8\%).

In conclusion, we have calculated the phase behavior of polydisperse
hard spheres, using accurate free energies for the fluid and solid
phases and solving exactly the resulting equilibrium conditions.
Fluid-solid coexistence has been identified for fluids with
polydispersities up to $\delta=14\%$. This shows clearly that the
experimentally observed suppression of crystallization above
$\delta=12\%$ is a non-equilibrium effect, probably caused by
increased nucleation barriers at large $\delta$~\cite{AueFre01}.  For
the solid, a terminal polydispersity remains well-defined as the
maximal value beyond which instability to phase separation sets in;
for triangular diameter distributions this turns out to be $\delta_t
\approx 7\%$. Instead of the re-entrant melting predicted in an
approximate treatment of fractionation effects~\cite{BarWar99}, we
find that sufficiently polydisperse solids split into two fractionated
solids on compression. At higher volume fractions and
polydispersities, multiple solids can coexist; coexistence of a fluid
with several solids appears as a new feature. Fractionation effects
are nontrivial, with solids splitting the diameter range unevenly
among them and coexisting fluids sometimes having larger
polydispersities than the parent.

Overall, our calculated phase diagram unites, clarifies and extends the
previous separate predictions of polydispersity effects on fluid-solid
coexistence and solid-solid fractionation. Numerical simulations may
offer the best avenue for testing our predictions but will need to be
carried out at fixed parent size distribution~\cite{WilSol02} to
detect the complex fractionation phenomena we find. For the future it
would be exciting to unify our predictions with those for fluid-fluid
demixing, but this will be very challenging since the latter only
occurs at polydispersities $\delta$ of order
$100\%$~\cite{Warren99,Cuesta99}, far outside the range studied here.


It is a pleasure to thank Paul Bartlett for providing his code for the
solid free energy. Financial support through EPSRC grant GR/R52121/01 is
acknowledged.


\bibliography{references}
\bibliographystyle{prsty}

\end{document}